\documentstyle[sprocl]{article}

\input{psfig}
\newcommand{\Pom}{{I\!\!P}}

\def\be{\begin{equation}}
\def\ee{\end{equation}}
\def\bea{\begin{eqnarray}}
\def\eea{\end{eqnarray}}

\begin{document}

\title{BFKL SIGNATURES AT A LINEAR COLLIDER}

\author{ C. ROYON}

\address{CEA, DAPNIA, Service de Physique des Particules, \\
Centre d'Etudes de Saclay, France}


\maketitle\abstracts{
The BFKL dynamics can be successfully tested at the 
future high energy $e^+e^-$ linear collider. The total $\gamma^*\gamma^*$
cross-section is calculated in the
Leading Log QCD dipole picture of BFKL  dynamics, and compared
with the one from 2-gluon exchange.
The rapidity dependence of the cross-section remains
a powerful tool to increase the ratio 
between the BFKL and the 2-gluon cross-sections and is more
sensitive to BFKL effects, even in the presence of higher
order corrections. Other potential signals like diffractive $J/ \Psi$
and 'forward' jet productions will also be discussed.
}

\vspace{-0.5cm}
\section{$\gamma^* \gamma^*$ total cross-section}
\subsection{Differential cross-sections}
Here, we want to calculate the total $\gamma^*\gamma^*$ cross-section derived in the 
Leading Log QCD dipole picture of BFKL dynamics. This could be a good test
of the BFKL equation which can be performed at e$^+$-e$^-$ colliders (LEP or
linear collider LC). The advantage of this process is 
that it is a process. which does not involve non perturbative couplings.

In this study, we compare the 2-gluon and the BFKL cross-sections. Defining 
$y_1$ (resp. $y_2$), and $Q_1^2$ (resp. $Q_2^2$) to be the rapidities and
the squared transfered energies for both virtual photons, one gets
\begin{eqnarray}
\label{eeBFKLa}
&&\hspace{-1.5cm}d\sigma_{e^+e^-}
 (Q_1^2,Q_2^2;y_1,y_2) = \frac{4}{9} \left(\frac{\alpha_{e.m}^2}{16}\right)^2  \, 
 \alpha_s^2 \, \pi^2 \sqrt{\pi} \,
\frac{d Q_1^2}{Q_1^2} \frac{d Q_2^2}{Q_2^2} \frac{d y_1}{y_1} 
\frac{d y_2}{y_2}
\frac{1}{Q_1 \, Q_2}
\, \nonumber \\
&&\frac{e^{\displaystyle \frac{4\alpha_s N_c}{\pi} Y \ln 2 }}{\sqrt{\frac{14
\alpha_s N_c}{\pi} Y \zeta(3)}}  
\times \, e^{\displaystyle -\frac{\ln^2 \frac{Q_1^2}{Q_2^2}}{\frac{56 
\alpha_s N_c}{\pi} Y \zeta(3)}}
 \left[2 l_1  +
9 t_1 \right] \, \left[2 l_2  +
9 t_2 \right] \,,
\end{eqnarray}
for the BFKL-LO cross-section, where
$t_1 = \frac{1 + (1-y_1)^2}{2}, \quad l_1=1-y_1$, and an analogous definition
for $t_2$ and $l_2$, 
and  
$Y=\ln \frac{s y_1 y_2 }{\sqrt{Q^2_1 Q^2_2}}$.
The 2-gluon cross-section has been calculated exactly within the high energy
approximation (NNNLO calculation) and reads
\begin{eqnarray}
\label{2g}
&~& d\sigma_{e^+e^-}
 (Q_1^2,Q_2^2;y_1,y_2) = \frac{d Q_1^2}{Q_1^2} \frac{d Q_2^2}{Q_2^2} 
\frac{d y_1}{y_1} \frac{d y_2}{y_2} 
 \, \frac{64 (\alpha_{e.m}^2 \alpha_s)^2}{243 \pi^3} \frac{1}{Q_1^2} 
\nonumber \\
&~& \left[ t_1 t_2 \ln^3 \frac{Q_1^2}{Q_2^2} + \left( 7 t_1 t_2 
+ 3 t_1 l_2 + 3 t_2 l_1 \right) \ln^2 \frac{Q_1^2}{Q_2^2} \right. \nonumber
\\ &~&  \left.
+ \left( ( \frac{119}{2} - 2 \pi^2 ) t_1 t_2
+ 5 (t_1 l_2 +t_2 l_1) + 6 l_1 l_2 \right) \ln \frac{Q_1^2}{Q_2^2} \right.
\nonumber
\\ &~& \left.
+ \left( \frac{1063}{9} - \frac{14}{3} \pi^2 \right) t_1 t_2
+ (46 - 2 \pi^2) (t_1 l_2 + t_2 l_1) - 4 l_1 l_2 \right] \,.
\end{eqnarray} 

Figure 1 shows the differential cross-sections in the
BFKL, DGLAP Double Leading Logarithm (DLL) and
2-gluon approximation,
as a function of $\ln Q^{2}_{1}/Q^{2}_{2}$ and for three values of $Y$.
The cross-sections on the left hand side
are calculated using  the exact unintegrated formulae,
for respectively the BFKL, DGLAP (in the double
Leading Log approximation) and 2-gluon exchange cross-sections.
Also the phenomenological HO-BFKL cross-sections, as detailed
in section 1.2, are given.

We note that the 2-gluon cross-section is almost always
dominating the DGLAP one in the Double Leading Log approximation.
The saddle point approximation turns out to be  a very good approximation
to an accuracy better than 10\%
for the BFKL cross-section and is not displayed in the figure.
We  note that the difference between the BFKL and 2-gluon
cross-sections increase with $Y$.
\par
On the right side of Figure 1,
curves for the exact LO and saddle-point
DGLAP calculations are
shown, as well as the full NNNLO (eq. \ref{2g}) result and the LO (eq. \ref{2g},
$\ln^3 Q^2_1/Q^2_2 $ term only)
result for the 2-gluon cross-section.
Unlike for the BFKL calculation, for the DGLAP
case the saddle-point approximation appears to be in worse agreement
with the exact calculation, and overestimates the cross-section by one
order of magnitude, which is due to the fact that
we are far away from the asymptotic regime. The comparison between the
DGLAP-DLL and the 2-gluon cross-section in the LO approximation shows that
 both cross-sections are similar when
$Q_1$ and $Q_2$ are not too different (the dashed line describes the
value $Q_1^2/Q_2^2=2$), so
precisely in the kinematical domain where the BFKL cross-section is
expected to dominate. However,
when $Q_1^2/Q_2^2$ is further away from one, the LO 2-gluon cross-section
is lower than the DGLAP one, especially at large $Y$. This suggests that
the 2-gluon cross-section could be a good approximation of the
DGLAP one if both are calculated at NNNLO and restricted
to the region where
$Q_1^2/Q_2^2$ is close to one. In this paper we will use the
exact NNNLO 2-gluon cross-section in the following to evaluate the
effect of the  non-BFKL background,
 since the 2-gluon term appears to constitute
 the dominant part of the DGLAP cross-section in the
region $0.5 < Q_1^2/Q_2^2 < 2$.

\subsection{Integrated cross-sections}

Results based on the calculations developed above 
will be given for 
 a future Linear Collider 
(500 centre-of-mass energy).
$\gamma^* \gamma^*$ interactions are selected at $e^+e^-$ colliders by detecting the 
scattered electrons, which leave the beampipe, in forward
calorimeters. For the LC it has been argued~\cite{brl}
that angles as low as 20 mrad should be reached. Presently
angles down to 40 mrad are foreseen to be instrumented for a generic
detector at the LC. 

Let us first specify the
region of validity for the parameters controlling
the basic assumptions made in the previous chapter. The main constraints
are required by the validity of the perturbative calculations.
The ``perturbative'' constraints are imposed by considering only photon 
virtualities $Q^{2}_{1}$, 
$Q^{2}_{2}$ high enough so that the scale $\mu^{2}$ in $\alpha_S$ 
is greater than 3 GeV$^{2}$. $\mu^2$ is defined 
using the Brodsky Lepage Mackenzie (BLM) scheme \cite{blm}, 
$\mu^2=\exp(- \frac{5}{3}) \sqrt{Q_1^2 Q_2^2}$.
In this case 
$\alpha_S$ remains always 
small enough such that the perturbative calculation is valid. 
In order that gluon contributions dominates the QED one, $Y$ 
is required to stay
larger than $\ln(\kappa)$ with $\kappa = 100.$ (see Ref. \cite{blm} for 
discussion).
Furthermore, in order to suppress DGLAP evolution, while maintaining
BFKL evolution  will constrain  $0.5 < Q_1^2/ Q_2^2 < 2$
for all nominal calculations.  In this paper we will use the
exact NNNLO 2-gluon cross-section in the following to evaluate the 
effect of the  non-BFKL background,
 since the 2-gluon term appears to constitute 
 the dominant part of the DGLAP cross-section in the 
region $0.5 < Q_1^2/Q_2^2 < 2$. 

We will not discuss here all the phenomenological results, and some detail can
be found in \cite{gamma}, as well as the detailed calculations.
We first study the effect of increasing the LC detector acceptance 
for electrons scattered under small angles and the ratio of the 2-gluon 
and the BFKL-LO cross-sections increase by more a factor 3 if the tagging 
angle varies between 40 and 20 mrad. The effect of lowering the tagging energy
is smaller. An important issue on the BFKL cross-section is the importance
of the NLO corrections and we adopt a phenomenological approach to estimate the effects
of higher orders. First, at Leading Order, the rapidity $Y$ is not 
uniquely defined, and we can add an additive constant to $Y$. 
A second effect of NLO corrections is to lower the 
value of the so called Lipatov exponent in formula
\ref{eeBFKLa}. In the $F_2$ fit described in Ref. \cite{ourpap}, the value of the 
Lipatov exponent $\alpha_\Pom$
was fitted and found to be 1.282, which gives an effective value of $\alpha_s$
of about 0.11. The same idea can be applied phenomenologically for the $\gamma^* 
\gamma^* $ cross-section. For this purpose, we modify the scale in $\alpha_S$ so that
the effective value of $\alpha_S$ for $Q_1^2=Q_2^2=25$ GeV$^2$ is about
$0.11$. 
Finally, the results of
the BFKL and 2-gluon cross-sections are given in Table \ref{1.fin} if we
assume both effects. 
The ratio
BFKL to 2-gluon cross-sections is reduced to 2.3 if both effects are
taken into account together. In the same table, we also give these effects for
LEP with the 
nominal selection and at the 
LC with a detector with increased angular acceptance. 

Another idea to establish the BFKL effects in data is to study the 
energy or $Y$ dependence of the cross-sections, rather than the comparison with 
total cross-sections itself. To illustrate this point,
we calculated the BFKL-NLO and the 2-gluon cross-sections, as well as
their ratio, for given cuts on rapidity $Y$ (see table 
\ref{fin.fin}). We note that we can reach
up to a factor 5 difference ($Y \ge 8.5$) keeping a cross-section
measurable at LC. The cut $Y \ge$ 9. would give a cross-section hardly
measurable at LC, even with the high luminosity possible at this
collider.

\begin{table}
 \begin{center}
\begin{tabular}{|c||c||c|c|c||c|} \hline
   & BFKL$_{LO}$     & BFKL$_{NLO}$ & 2-gluon     & ratio \\ \hline \hline
 LEP        & 0.57   & 3.1E-2       & 1.35E-2  & 2.3     \\
 LC 40 mrad & 6.2E-2 & 6.2E-3       & 2.64E-3  & 2.3     \\
 LC 20 mrad & 3.3    & 0.11         & 3.97E-2  & 2.8     \\ \hline
\end{tabular}
\caption{Final cross-sections (pb), for selections described in the text.}
\label{1.fin}
 \end{center}
\end{table}

\begin{table}
 \begin{center}
\begin{tabular}{|c||c||c|c|c||c|} \hline
$Y$ cut   & BFKL$_{NLO}$     &  2-gluon     & ratio \\ \hline
\hline
 no cut & 1.1E-2 & 3.97E-2 & 2.8 \\
 $Y \ge $ 6. & 5.34E-2 & 1.63E-2 & 3.3 \\
 $Y \ge $ 7. & 2.54E-2 & 6.58E-3 & 3.9 \\
 $Y \ge $ 8. & 6.65E-3 & 1.43E-3 & 4.7 \\
 $Y \ge $ 8.5 & 1.67E-3 & 3.25E-4 & 5.1 \\
 $Y \ge$ 9. & 5.36E-5 & 9.25E-6 & 5.8 \\ \hline
\end{tabular}
\caption{Final cross-sections (pb), for selections described in the
text, after different cuts on $Y$}
\label{fin.fin}
 \end{center}
\end{table}

\begin{figure}
\psfig{figure=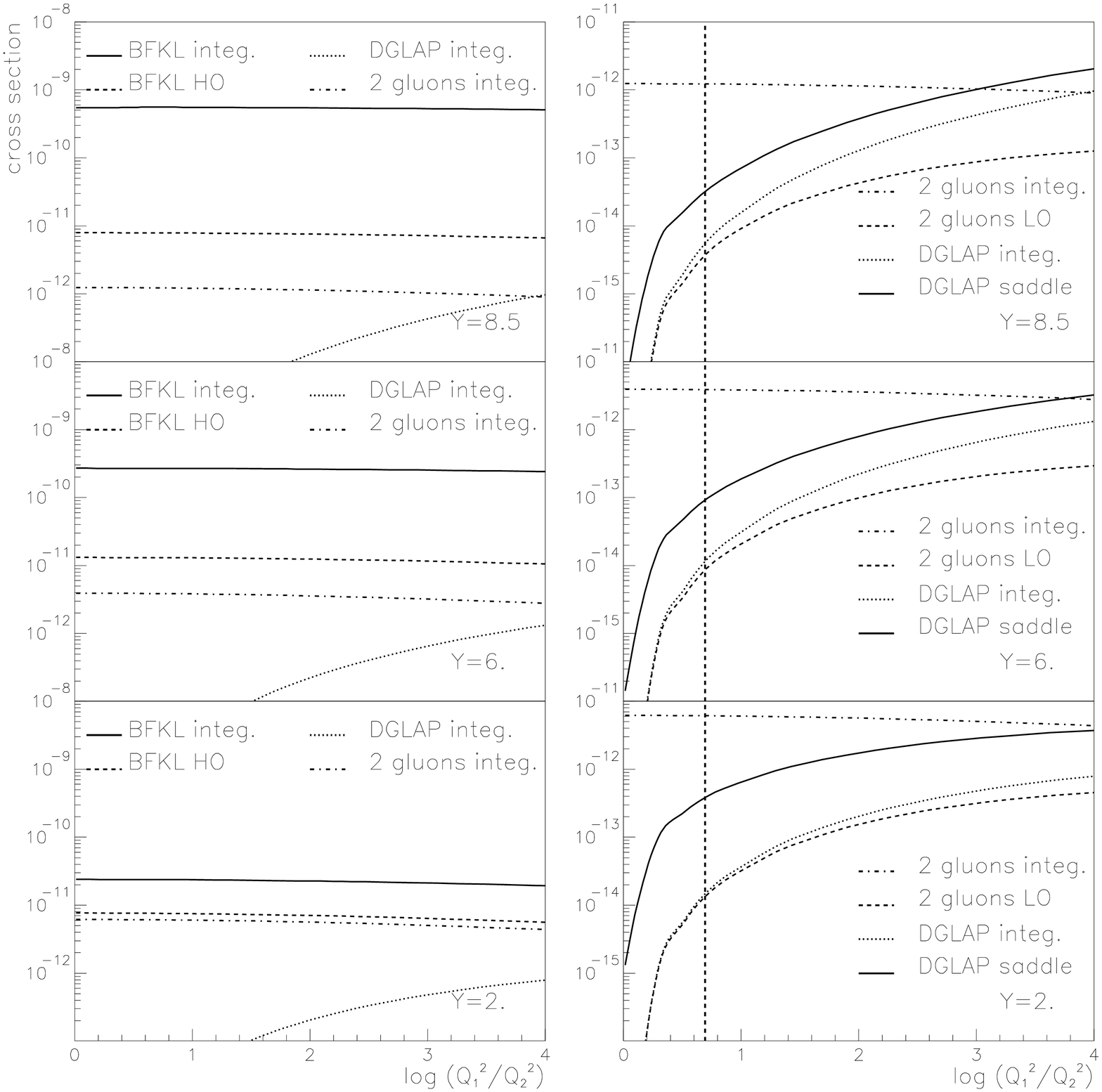,height=5.in}
\caption{Differential cross-sections for 
 different values of $Y$ (see text).}
\end{figure}

\vspace{-0.5cm}
\section{Diffractive $J/\Psi$ production as a probe of the QCD pomeron}
Another probe of the QCD pomeron at LC which has been proposed is
the double diffractive production of
$J/\Psi$ in $\gamma \gamma$ collisions i.e. the process $\gamma
\gamma \rightarrow J/\Psi  J/\Psi$ \cite{Kwiecinski}.
It should be noted that both sides
of the diagram are characterized by the same (hard) scale
provided in this case   by the relatively
large  charm quark mass. This process has also the advantage that
its cross-section  can
be almost entirely calculated perturbatively. The only non-perturbative
element  is a  parameter determined by the $J/\Psi$ light cone
wave function
which can however be obtained from the  measurement of the leptonic
width $\Gamma _{J/\Psi \rightarrow l^+l^-}$ of the $J/\Psi$. 
The BFKL equation is solved in the non-forward
configuration taking into account dominant non-leading
effects which come from the requirement that the virtuality of the
exchanged gluons along the
gluon ladder is controlled by their transverse momentum squared.

The cross-section exhibits an
approximate $(W^2)^{2\lambda}$ dependence. The parameter $\lambda$
slowly increases  with increasing energy $W$ and changes from
$\lambda \approx 0.23$ at $W=20$ GeV to $\lambda \approx 0.28$ at
$W=500$ GeV i.e. within the energy range which is relevant for
possible LC measurements.

The BFKL effects significantly affect
the $t$-dependence of the differential cross-section leading to steeper
$t$-dependence than that generated by the Born term. Possible
energy dependence of the diffractive slope is found to be very
weak. 

The 2-gluon and BFKL cross-sections after the consistency constraints are still
quite different at LC and it would be possible to see clearly some BFKL
enhancement of the cross-section. At at center of mass energy of 200 GeV
(resp. 500 GeV), the 2-gluon and BFKL cross-sections are 0.03 and 0.13 pb
(resp. 0.11 and 0.65 pb). Going to  the $e \gamma$ option instead of the 
$e^+ e^-$ one increases sensitively the BFKL cross-section. At a center of mass
energy of 400 GeV, the BFKL cross-section is about 6.5 pb whereas the 2-gluon
one is about 0.8 leading to a clear distinction between them.

Unfortunately, these cross-sections are purely theoretical as they do not
take into account the angular acceptance of the detectors. If it is
possible to detect the decayed muon coming from the $J/ \Psi$ down to
5 degrees (resp. 15 degrees), it is possible to detect 25\% (resp. 4\%)
of the cross-sections discussed above inside the acceptance of the detector.
Thus, about 50, 500 and 5000 such events are expected at LC in the
$ee$, $e \gamma$, and $\gamma \gamma$ options for 100 fb$^{-1}$.
To be able to study this kind of processes, high luminosity as well as a 
good coverage of the muon detectors in the forward and backward part down
to 5 degrees is necessary.

In the same spirit, another measurement based on diffractive photon production
in $\gamma \gamma$ interactions has been proposed in \cite{Forshaw}. The typical
cross-section is of the order of 1 pb at LC for $t$ between 0 and 10$^{-4}$.
The study within the detector acceptance is under way.

\vspace{-0.5cm}
\section{"Forward jets" production}
The idea of looking for "forward jets" at LC follows the motivation of the
search for effects of BFKL-like parton shower evolutions performed at HERA
\cite{fwdjets}. The method proposed originally by A.Mueller \cite{Muellerb}
for looking for BFKL effects at HERA starts from the idea that the BFKL
evolution predicts a different momenta ordering in the parton cascade 
compared to the DGLAP one. The DGLAP evolution predicts a strong ordering
in the parton transverse momenta $k_T$ whereas the BFKL one relaxes this
ordering. Thus the BFKL evolution predicts additional contributions to the
hadronic final state coming from partons with large transverse momenta, and
high transverse momenta partons going forward in the HERA frame. The same
idea can easily be applied to LC if one considers on one side a virtual 
photon of virtuality $Q^2$ interacting with a resolved 
photon \cite{Albert}. 
If one requires
the jet close in rapidity to the resolved photon to have a $k_T^2$ close
to $Q^2$, one enhances the BFKL evolution. The DGLAP cross-section will be
strongly suppressed because of the $k_T$ ordering whereas the BFKL evolution
will predict a non vanishing cross-section.

A detailled similation of this effect has been performed for LC \cite{Albert}.
To be able to tag the "forward jet", cuts on the jet angle larger than
5 degrees, and on the jet transverse energy greater than 4 GeV have been
performed. To enforce the BFKL cross-section compared to the DGLAP one,
a cut on the ratio of the tranverse energy of the jet over $Q^2$ has been
added ($0.5 < k_T^2 / Q^2 < 2$). This leads to a LO-BFKL over Born
cross-section ratio varying between 0.05 and 0.1. It should also be noticed
that the $e \gamma$ collider provides a 10-time higher cross-section than
the $ee$ option. In figure 2 is displayed the result for the Born and 
LO-BFKL cross-section as well as their ratio for an $ee$ (full line) and
$e \gamma$ (dashed line) machine.

\begin{figure}
\begin{center}
\psfig{figure=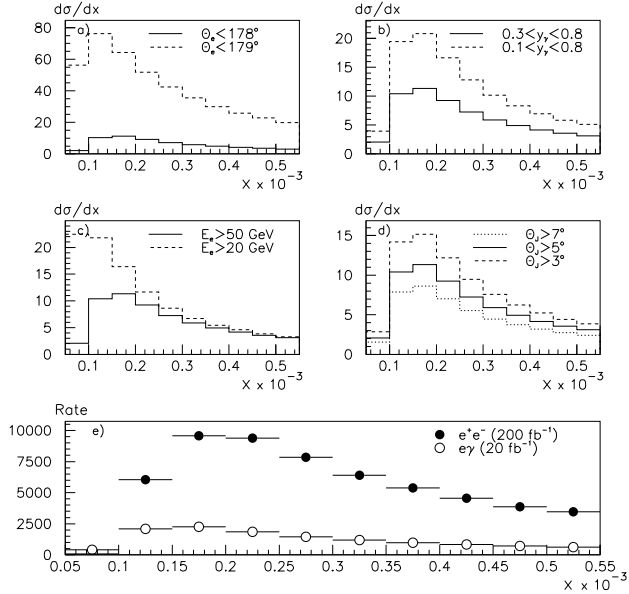,height=3.5in}
\end{center}
\caption{Forward jet cross-sections for an $ee$ and an $e \gamma$ collider for
different cuts on the electron energy and angle, and on the jet energy and
angle.
}
\end{figure}

\vspace{-0.5cm}
\section{Conclusion}

We first discussed the difference between the 2-gluon and BFKL
$\gamma^* \gamma^*$ cross-sections 
both at LEP and LC. The LO BFKL cross-section is much larger than the 2-gluon cross-section.
Unfortunately, the higher order  corrections 
of the BFKL equation (which we estimated phenomenologically)
are large, and the 2-gluon and BFKL-NLO cross-sections ratios are 
reduced to a factor two to four.
The $Y$ dependence of the cross-section remains
a powerful tool to increase this ratio and is more 
sensitive to BFKL effects, even in the presence of large higher 
order corrections. The uncertainty on the BFKL cross-section after
higher order corrections is still quite large. 
We thus think that the measurement performed at LEP
or at LC should be compared to the precise calculation of the 2-gluon
cross-section after the kinematical cuts described in this paper, and the
difference can be interpreted as BFKL effects. A fit of these
cross-sections will then be a way to determine the BFKL pomeron intercept
after higher order corrections. A possible meaurement at LC would then be
of great importance provided it is possible to tag electrons at low 
scattering angles.

The second measurement described in this paper is the double diffractive production of
$J/\Psi$ in $\gamma \gamma$ collisions. The measurement of its cross-section
(and more specifically of its $t$ dependence) allows a clear distinction
between the DGLAP and BFKL evolutions. A good coverage at low angle of the
muon detection is then needed. If this is fulfilled, a clear BFKL signal
could be shown at LC with this measurement.

One of the golden channels for BFKL searches at HERA, i.e.
forward jet production can also been used at LC provided tagging of jets
at low angle is feasible. The predicted ratio between the BFKL-LO and 2-gluon
cross-sections is then found to be between 0..05 and 0.1.
The effects of higher order corrections to the 
BFKL equation still need to be studied for this process.

\end{document}